\documentclass[nocompress]{spie}  

\usepackage{amsmath,amsfonts,amssymb}
\usepackage{graphicx}
\usepackage[colorlinks=true, allcolors=blue]{hyperref}
\usepackage{soul}
\usepackage{cancel}

\usepackage{multirow}

\title{AIROPA II: Modeling Instrumental Aberrations for Off-Axis Point Spread Functions in Adaptive Optics}

\author[a]{Anna Ciurlo} 
\author[b]{Paolo Turri} 
\author[a,c]{Gunther Witzel} 
\author[d]{Jessica R. Lu} 
\author[a]{Tuan Do} 
\author[a,e]{Breann N. Sitarski}
\author[a]{Michael P. Fitzgerald} 
\author[a]{Andrea M. Ghez} 
\author[f]{Carlos Alvarez}
\author[d]{Sean K. Terry} 
\author[f]{Greg Doppmann}
\author[f]{James E. Lyke} 
\author[f]{Sam Ragland} 
\author[f]{Randall Campbell} 
\author[g]{Keith Matthews}

\affil[a]{University of California, Los Angeles, Department of Physics and Astronomy}
\affil[b]{University of British Columbia}
\affil[c]{Max-Planck-Institut fuur Radioastronomie}
\affil[d]{University of California, Berkeley, Astronomy Department}
\affil[e]{Giant Magellan Telescope}
\affil{University of California, Los Angeles, Department of Physics and Astronomy}
\affil[f]{W.~M. Keck Observatory}
\affil[g]{California Institute of Technology} 
 
\authorinfo{Send correspondence to A. Ciurlo. E-mail: ciurlo@astro.ucla.edu}

\begin{document} 
 
\maketitle

\begin{abstract}
Images obtained with single-conjugate adaptive optics (AO) show spatial variation of the point spread function (PSF) due to both atmospheric anisoplanatism and instrumental aberrations.
The poor knowledge of the PSF across the field of view strongly impacts the ability to take full advantage of AO capabilities.
The AIROPA project aims to model these PSF variations for the NIRC2 imager at the Keck Observatory. 
Here, we present the characterization of the instrumental phase aberrations over the entire NIRC2 field of view and we present a new metric for quantifying the quality of the calibration, the fraction of variance unexplained (FVU).
We used phase diversity measurements obtained on an artificial light source to characterize the variation of the aberrations across the field of view and their evolution with time.
We find that there is a daily variation of the wavefront error (RMS of the residuals is 94~nm) common to the whole detector, but the differential aberrations across the field of view are very stable (RMS of the residuals between different epochs is 59~nm).
This means that instrumental calibrations need to be monitored often only at the center of the detector, and the much more time-consuming variations across the field of view can be characterized less frequently (most likely when hardware upgrades happen).
Furthermore, we tested AIROPA's instrumental model through real data of the fiber images on the detector.
We find that modeling the PSF variations across the field of view improves the FVU metric by 60\% and reduces the detection of fake sources by 70\%.
\end{abstract}


\section{Introduction}

Adaptive optics (AO) has revolutionized ground-based astronomy through its delivery of diffraction limited images at infrared wavelengths. 
Despite its success, we have not yet fully capitalized on
AO capabilities, primarily due to some fundamental limitations of first-generation AO systems. 
The most prevalent type of AO system on telescopes today is a single-conjugate AO system that corrects for the integrated turbulence of the atmosphere as measured by a single guide star, typically at the center of the field of view.
Astronomical objects that are extended over the field show a degradation in the AO correction further from the guide star due to the differences in the columns of atmosphere the light from different directions passes through. 
This anisoplanatism leads to a point spread function (PSF) that varies both with time and position in the field, which is difficult to characterize \cite{Schodel:2010}.

In addition to anisoplanatism introduced by the Earth's atmosphere, the telescope and instrument itself can introduce field-dependent aberrations that change the PSF. 
These aberrations are a major contribution to PSF variability throughout the field that limits many scientific applications. 
These instrumental aberrations are particularly problematic as they can be highly asymmetric and introduce astrometric offsets. 
Extracted astrometry, photometry, and morphology are impacted by this lack of knowledge of the PSF variability over the field. 
In particular, astrometric measurements in crowded fields such as the Galactic Center are dependent on precise knowledge of the PSF shape at each position in the field of view and the astrometric reference within each PSF. 

In order to overcome some of these limitations, we have developed a new software package, AIROPA (Anisoplanatic and Instrumental Reconstruction of Off-Axis PSFs for AO), dedicated to modeling the spatial variation of the PSF in single-conjugate AO images \cite{Witzel:2016}. 

This effort is based on a hybrid approach of extracting knowledge of the PSF at the pointing position from the imaging data, and then using a convolution kernel for each position in the field to model the local PSF \cite{Britton:2006,Fitzgerald:2012}: AIROPA uses an empirical on-axis PSF and through its atmospheric and instrumental models predicts its variations across the field of view.
The atmospheric model of AIROPA builds on a study for the Palomar telescope \cite{Britton:2006}, extending it from NGS to the LGS case. The instrumental model is provided by using phase maps that characterize instrumental aberrations.
In practice, AIROPA uses the product of the optical transfer function (OTF) corresponding to each term (empirical, atmospheric and instrumental PSFs) to estimate the PSF across the field of view.  

AIROPA is described in a series of papers: An overview is given in Ref.~\citenum{Witzel:2016}.
In this paper, we concentrate on the characterization of the instrumental aberrations. 
Testing with simulations and on-sky images are presented in Ref.~\citenum{Turri22}, and on-sky testing in a wide range of atmospheric conditions is discussed in Ref.~\citenum{Terry:inprep}.

AIROPA has been developed for the NIRC2 imager (PI: K. Matthews) at the W.~M.~Keck Observatory, but the framework is built to be easily adaptable to a range of other instruments \cite{Ciurlo:2018, Do:2018}.
NIRC2 is fed by the AO system of the Keck II telescope. 
NIRC2 has a 1024$\times$1024 pixels InSb Aladdin-3 detector and a $\sim$10~mas pixel scale for its narrow field camera.
This instrument has been used extensively for astrometric programs such as monitoring stellar orbits around the supermassive black hole at the Galactic Center \cite{Ghez:2008, Boehle:2016, Do:2019}; thus, PSF characterization efforts are particularly useful for NIRC2.

The AO system at Keck is a single-conjugate system (hence it corrects the total integrated atmospheric turbulence \cite{Wizinowich:2006,vanDam:2006}). 
Keck AO wave-front sensing is performed on either a natural guide star or a laser guide star using a Shack-Hartmann wave-front sensor that operates at optical wavelengths. 
\footnote{
Keck AO has been recently updated with a near-infrared pyramid wavefront sensor\cite{Bond20}, but the work here focuses on data taken in with the Shack-Hartman sensor, used in the past 20 years.
}

During routine AO operations, afternoon image sharpening is performed using a fiber source in the focal plane near the entrance of the AO system \cite{vanDam:2006}. 
The fiber is imaged through the AO system and onto the center of the science detector (or more precisely, the pointing origin of the telescope near the center) in the NIRC2 camera. 
Static aberrations from the AO system and NIRC2 camera are estimated from the fiber images using a phase diversity method. 
Phase diversity allows to measure the wavefront phase from intensity maps of in-focus (or near focus) point-like sources.
In this case the phase is extracted through a Gerchberg-Saxton method \cite{Gerchberg:1972}. 
Despite this optimization at the center, the optical system shows degradation of the PSF toward the edges of the field.
In particular, significant elongation is present in all four corners \cite{Gautam:2019}. 

To characterize the instrumental aberrations, we further calibrate the instrument after image sharpening: we adapt the image sharpening phase diversity procedure to estimate aberration phase maps at a grid of positions over the whole field of view. 
In Section~\ref{sec:dataset} we describe the data observations and reduction.
In Section~\ref{sec:analysis} we describe the retrieval of phase maps and we introduce performance benchmarks in PSF metrics.
In Section~\ref{sec:results} we describe the spatial and temporal variability of the phase maps and we test AIROPA instrumental model on observations of the fiber source. 
Our results and conclusions are summarized in Section~\ref{sec:conclusion}.

\section{Dataset: phase diversity}
\label{sec:dataset}

Phase diversity data are taken during the afternoon operations, using a calibration fiber that can be moved across the field of view and at different distances from the field-of-view center. 
All the data is taken with the following NIRC2 setup: narrow camera ($\sim$10~mas/pixel), FeII filter (with central wavelength 1.6455~$\mu$m and band-pass 0.0256~$\mu$m) and open pupil.
The pupil we use is circular in shape, with no central obstruction, and has a size slightly larger than the telescope's pupil. The different size should not affect the PSF-reconstruction process on sky because the phase maps are then masked by AIROPA using the correct telescope pupil shape, size, and rotation. 

We note that any wavefront change between the  Kp filter or H filter (commonly used for science) and the FeII filter will be field independent because they are situated at the image of the entrance pupil.
The fiber has its limitations: it’s not always very stable in brightness and it's not very uniformly lit. We address the brightness variation problem by tuning our exposure times so as to always have a constant number of peak counts. The fiber, albeit imperfect, seems to perform well for image sharpening and it’s thus likely good enough for phase diversity. Moreover in Section~\ref{subsec:prediction}, we show that through AIROPA we manage to retrieve very well the PSF of in-focus fiber images, further proving that the phase diversity data is good enough for our purposes.
First, the routine image sharpening is performed at the NIRC2 pointing origin near the center of the detector where the residual wavefront caused by aberrations is minimal. 
After image sharping we keep the correction on the deformable mirror constant for our subsequent measurements.
  
\begin{figure}[h]
\begin{center}
\includegraphics[width=14cm]{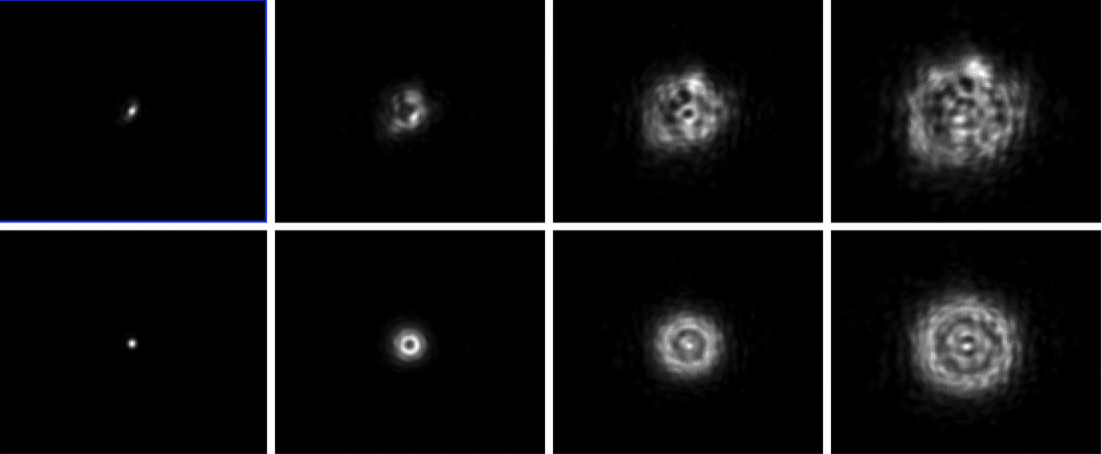}
\caption{\label{fig:focpos}
  From left to right: in-focus and out-of-focus (focus offsets -2,-4,-6 mm) fiber images in the top-right corner (top) and the center (bottom) of the NIRC2 detector. 
}
\end{center}
\end{figure}

To produce a phase map for a single NIRC2 detector position we take a set of three images (for bad pixel removal, see next section) of the fiber respectively at three different out-of-focus steps (-2,-4,-6 mm, z-positions of the fiber). 
We also collect, at each detector position, an image of the fiber in-focus for the testing of the code (see Section~\ref{subsec:prediction}). 
An example of in-focus and out-of-focus fiber images is reported in Figure~\ref{fig:focpos}.

Phase diversity data requires a high signal to noise ratio for the complex out-of-focus speckles from which phase information is derived. 
In particular, the luminosity of the fiber needs to be adapted for the most out-of-focus position because the fiber light is spread over a larger number of pixels.
This is done using different neutral density filters in front of the fiber source to dim (for in-focus images) or enhance (for the most out-of-focus images) its light.
The integration times are adjusted accordingly, to optimize exposure time, signal-to-noise and avoid saturation.
We adjust the observations settings to obtain a signal-to-noise ratio of 20.
We repeat this process for different positions across the detector to collect information on the filed-dependent instrumental aberrations.
Calibration data, such as background images with the fiber turned off and  detector flat field are recorded as well. 

\begin{figure}[h]
\begin{center}
\includegraphics[width=5cm]{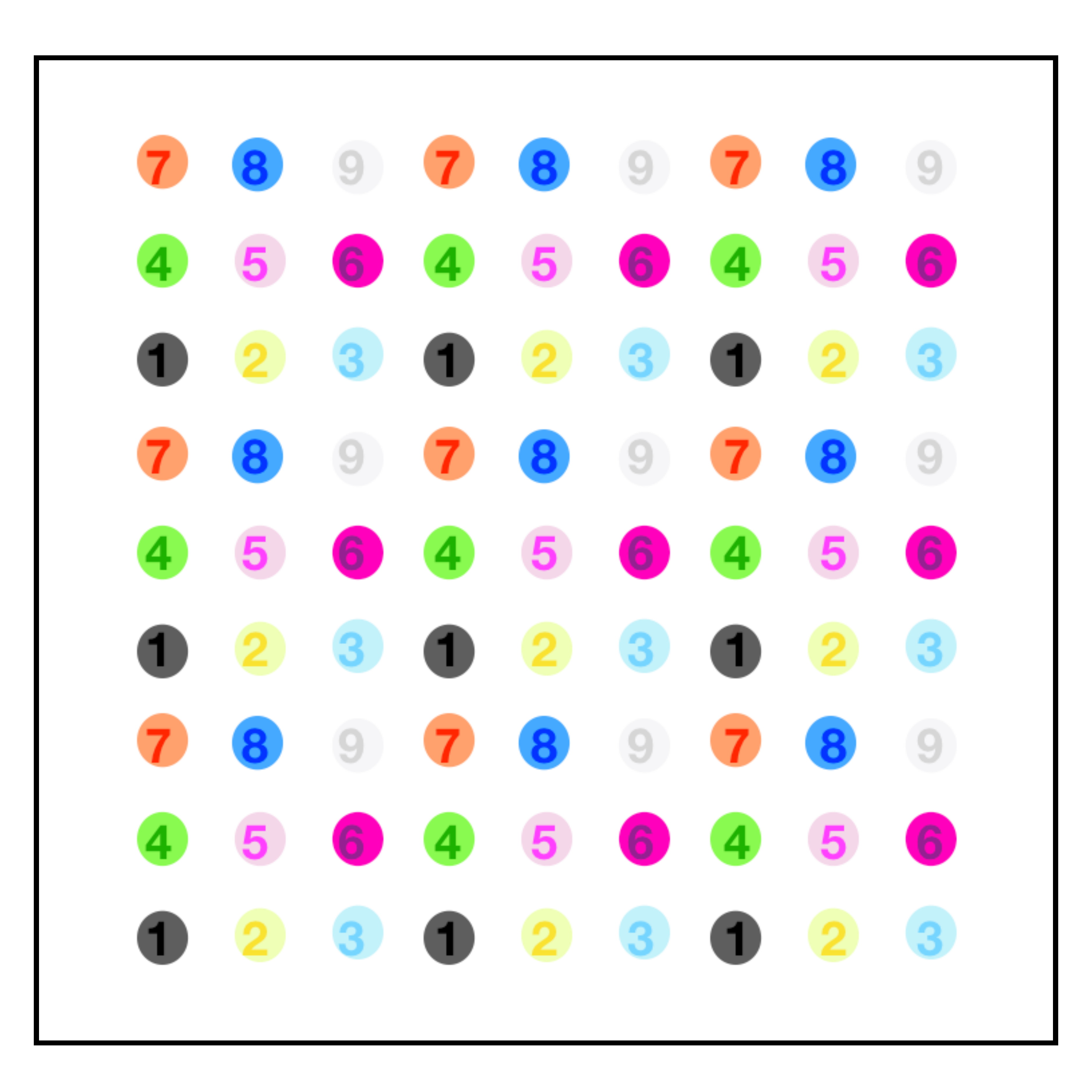}
\caption{\label{fig:fiber_data}
  Grid of positions on the NIRC2 detector where phase diversity data are taken (each grid is identified by a different color). 
  The out-most $\sim$100 pixels in x and y are not covered by this grid because of the dimensions of the fiber image at the most extreme fiber stage position (-6 mm).
}
\end{center}
\end{figure}

In order to facilitate a fast daytime measurement scheme, that can fit in the scheduled calibration and telescope operations, we implemented an efficient NIRC2 observing script that results in nine 3$\times$3 fiber grids, with three frames for each of the out-of-focus (-2,-4,-6 mm) and in-focus fiber position. 
This configuration allows us to have a quick snapshot of aberrations across the detector for each 3$\times$3 grid while producing a fine enough grid when combining all nine grids.
All nine grids can be taken in about 7 hours total and result in a final grid of 9$\times$9 phase maps over the NIRC2 narrow camera field of view (Fig.~\ref{fig:fiber_data}). 

We started with an exploratory phase where many positions were sampled between 2014 and 2015. 
In 2017 we began using a more standardized approach, observing a series of 3$\times$3 grids. 
All our observations so far are reported in Table~\ref{tab:obs} 
The typical integration time for the out-of focus images (adjusted for the varying fiber brightness) is 3 times the in-focus time.

\begin{table}[]
    \centering
    \begin{tabular}{|c|p{1.8cm}|c|p{.6cm}|c|c|}
\hline
Year & Epochs & Pos & Cen & TR & TR-Cen \\
\hline
Pre-2015 & 
121016 
121015 
120722 
131211
131104
131014
130822
130821
130508
130410
130307
130208
140607 
140513 &
145 & -- & -- & -- \\
\hline
2015 &
150514
150528
150610
150611
150615
150722
150809
150810
151102
151105 
& 145 
& 59 59 59 63 66 69 73 74 86 90 
& 205 & 196 \\
\hline
2017 & 170728 & 82 & 75 & 238 & 205 \\
\hline
2018 & 180901 & 19 & 77 & 216 & 200 \\
\hline
2019 & 190511 190603 & 82 & 52 56 & 204 & 200 \\
\hline
2021 & 211003 211013 & 82 & TBD TBD & TBD & TBD \\
\hline
    \end{tabular}
    \caption{
    Diversity data gathered for monitoring instrumental aberrations on NIRC2. For each year we report the actual observing dates (format YYMMDD), the total number of positions observed, the WFE in nm at the center ($Cen$), the top-right corner ($TR$) and the differential between top-right corner and center ($TR-Cen$). }
    \label{tab:obs}
\end{table}

\section{Analysis}
\label{sec:analysis}
\subsection{Phase maps retrieval}
\label{subsubsec:retr}
 
Removal of bad pixels is particularly important as the phase retrieval code struggles to converge in the presence of cosmic rays.
A bad pixels mask is created from the background images. 
The individual NIRC2 exposures are then background subtracted, flat-field and bad pixels corrected.
For each position, we take the median of the three images, which significantly reduces the number of residual bad pixels.
In the following analysis we only use calibrated, median images.
We remove remaining bad pixels from the frames using a simple routine that rejects extreme values and replaces them by interpolating over neighboring pixels.

For each detector position and z-position we create a 256$\times$256 pixel image that is centered using the in-focus frames. 
The 256$\times$256 cutout of the out-of-focus images needs to be correctly aligned to avoid introducing spurious structures in the final phase map. 
We do the centering onto the in-focus images in 2 steps: 1) A rough first estimate with a centroid routine applied where the signal-to-noise
exceeds 5, 2) a fine-tuned estimate on a cutout around the previously determined center using a 2D-Gaussian fit.
The in-focus center position is used to extract the out-of-focus cutouts. 
Any residual tilt present in the images is dominated by the position of the fiber stage with respect to the detector and it thus not relevant for our analysis. Therefore we adopt the same procedure employed for image sharpening at Keck and remove the tilt. The characterisation and correction of the distortion is a separate part of our analysis\cite{Yelda10,  Service16}.
 

The three resulting cut-out, out-of-focus images are fed to a Gerchberg–Saxton (GS) algorithm \cite{Gerchberg:1972} to retrieve the phase.
The GS algorithm is an iterative algorithm that allows to retrieve maps of the wavefront phase by measuring wavefront intensities at multiple optical planes.
This algorithm is widely used for aberration correction and a modified version of it \cite{Atcheson:2003,Vandam:2016} is routinely employed at Keck for the image sharpening procedure on NIRC2 and other instruments.
Typically, the RMS error of the phase map after imaging sharpening is $\sim$68~nm.

However, image sharpening is exclusively performed at (or near the) center of the detector, and it is used in an iterative way to optimize the shape of the deformable mirror to remove non-common path aberrations. 
In our case, we want to account for all the spatially variable instrumental aberrations and store the information to be later used in AIROPA.
Therefore, adapting the code routinely used at Keck required some fine-tuning of the parameters used in the algorithm.

In the standard image sharpening routine the code is run with a low number of iterations of the GS algorithm (typically 5). 
For image sharpening, only a rough first estimate is necessary to provide a first order correction to transmit to the deformable mirror.
The algorithm is then run again with the new deformable mirror shape and improved upon.
This process is repeated several times until the resulting phase map is flat.

\begin{figure}[h]
\begin{center}
\includegraphics[width=14cm]{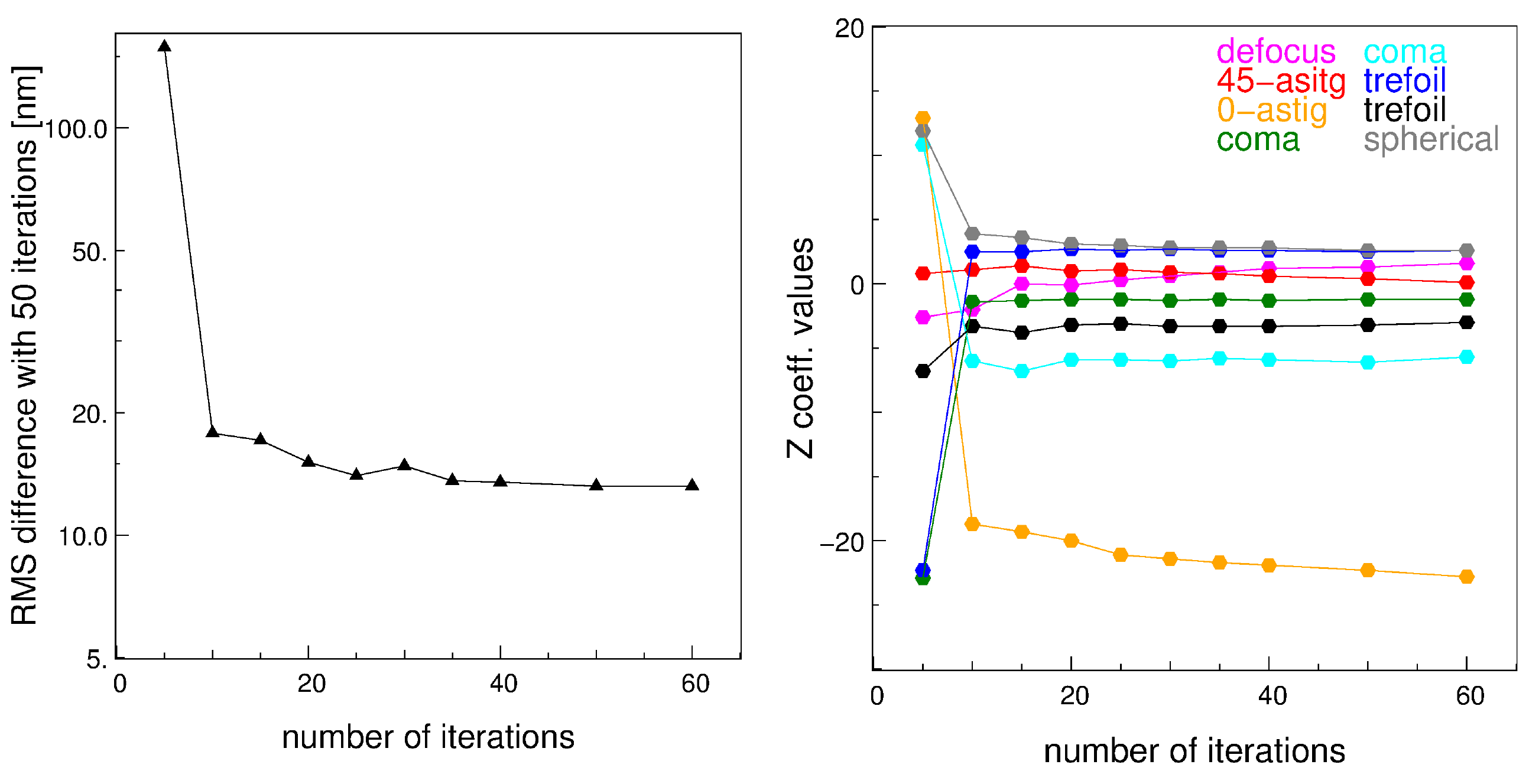}
\caption{
\label{fig:method_niter}
Left: Variations of the coefficients depending on the number of iterations of the GS algorithm (as printed by the phase diversity code). Right: RMS of the difference between a phase map obtained from a given number of iterations and 50 iterations phase map. 
}
\end{center}
\end{figure}

In our case, we need to obtain a precise estimate of the phase with just one run of the GS algorithm.
For a low number of iterations, especially for the positions further from the center, the procedure struggles to converge.
We tested the value of the Zernike coefficients obtained through the algorithm as a function of the number of iterations (Figure~\ref{fig:method_niter}) for one corner of the NIRC2 detector.
At 10 iterations the result became stable and coherent, while before that, the phase maps have poor quality and the coefficients are randomly distributed. 
The time the phase retrieval code requires to run is directly proportional to the number of iterations: 50 iterations requires about 10 min.
20 iterations would seem sufficient to have stable coefficients and faster integration times.
However, when running the code for several detector positions (our grid is composed of 81 positions) the algorithm does not properly converge. 
This most likely happens because the algorithm remains stuck on a local minimum. 
These cases can be identified when looking at the overall shape of the phase maps across the detector, as they look discontinuous with respect to the surrounding ones. 
The phase retrieval code involves a random procedure that adds noise to an initial guess phase map. 
The code can be re-run with a different seed and that typically fixes the issue.
However, it is time consuming to identify the non-converged phase maps and manually re-run that position in a procedure that is otherwise automatized.
We find that this problem can by avoided by using a higher number of iterations.
In summary, we find that the lowest number of iterations we can use to minimize the code running time and still have the code converge most of the times is 50.

We define algorithm noise as the RMS of the difference between two phase maps obtained from the same dataset but with different seeds. 
Typically, the RMS of residuals between these two phase maps is $\sim$13.9~nm. 

We constructed the phase maps from the median of 3 frames for each out-of-focus position. 
We compared this approach to building phase maps independently and then median-combining them. 

The average RMS for phase maps constructed from 3 frames is 17~nm, whereas when we construct 3 phase maps from each single frame set we obtain an RMS of 19 nm.
We conclude that the difference is small between the two approaches (and smaller than the algorithm noise). 
However, median-combining the frames is computationally faster and removes cosmic rays, thus we keep this approach.

\subsection{PSF Metrics}
\label{subsec:metrics}

To verify the quality of this model of the instrumental aberrations, as a first step we want to verify AIROPA's ability to reproduce the in-focus fiber images at different positions on the field of view.
In Ref.~\citenum{Turri22} and \citenum{Terry:inprep} we apply AIROPA to simulated and on-sky data. 
Before we can proceed, we need to establish clear metrics to measure AIROPA's performances.

In previous publication on this project \cite{Witzel:2016} we used to use fractional error (FE), the RMS of the residual over a 40$\times$40 pixel box divided by the PSF's average intensity over the same box:
\begin{equation}
FE = \frac{\sqrt{\sum_x \sum_y [PSF_{obs}(x, y) - PSF_{mod}(x,
    y)]^2}}{\sum_x \sum_y PSF_{obs}(x, y)}
\end{equation}
where $PSF_{mod}$ and $PSF_{obs}$ are the modeled and observed PSFs respectively. 
We note that the PSF is flux-normalized during the extraction phase in AIROPA. The position of the model PSF is not aligned to the observed one as we do not know the \textit{a priori} position of the fiber image.
While this yields some comparability between residuals of stars in the same on-sky frame, it cannot be interpreted as a typical residual amplitude in percent. 
In particular, this metric is very dependent on the particular shape of the PSF and it is not a valid comparison of the quality of PSF modeling across wavelengths, instruments, observing night with different quality of AO-correction, or even across modeling results from different phase maps. 

We find a much more appropriate metric by subtracting the mean over the full PSF support size of 40x40 pixels ($\bar{PSF}_{obs}$):
\begin{equation}
FE2 = \sqrt{\frac{\sum_x \sum_y [PSF_{obs}(x, y) - PSF_{mod}(x,
    y)]^2}{\sum_x \sum_y [PSF_{obs}(x, y) - \bar{PSF}_{obs}]^2}}
\end{equation}
This quantity can be interpreted as a measure of the typical (average) pixel-wise deviation of the PSFs, in percent of the PSF intensity. 
The square of the fractional error above (FE2) is a statistical quantification on how well the the variance in the data is reproduced by the model, the so called fraction of unexplained variance or FVU:
\begin{equation}
FVU = \frac{\sum_x \sum_y [PSF_{obs}(x, y) - PSF_{mod}(x,
  y)]^2}{\sum_x \sum_y [PSF_{obs} - \bar{PSF}_{obs}]^2}
\end{equation}
The FVU is closely related to the so-called coefficient of determination, R$^2$, which is a well characterized statistical tool \cite{coeff}.

\section{Results}
\label{sec:results}
\subsection{Phase maps variability}
\label{subsec:variability}

\begin{figure}[h]
\begin{center}
\includegraphics[width=10cm]{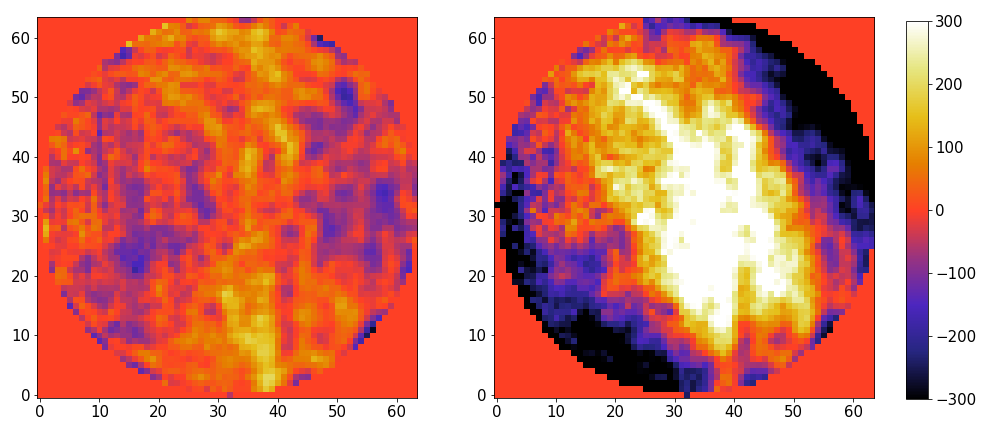}
\caption{\label{fig:center_to_edge}
  Central and corner phase maps for the NIRC2 detector shown with the same scale, in nm. In this example the central and edge phase maps have a RMS of 76 and 216~nm respectively.
}
\end{center}
\end{figure}

In Figure~\ref{fig:center_to_edge}, we show an example of phase maps at the center and in the corner of the detector obtained in 2018.
Table \ref{tab:obs} reports the wavefront error at the center of the detector and in the top-right corner for every year we took data.
The RMS of the phase maps at the corner are the highest across the field of view, typically over 200~nm (as expected, as they represent the point further out from where image sharpening is performed).
There are also differences among the corners themselves, typically the RMS of the residuals of opposite corners is about 125~nm.

\begin{figure}[h]
\begin{center}
    \includegraphics[width=15cm]{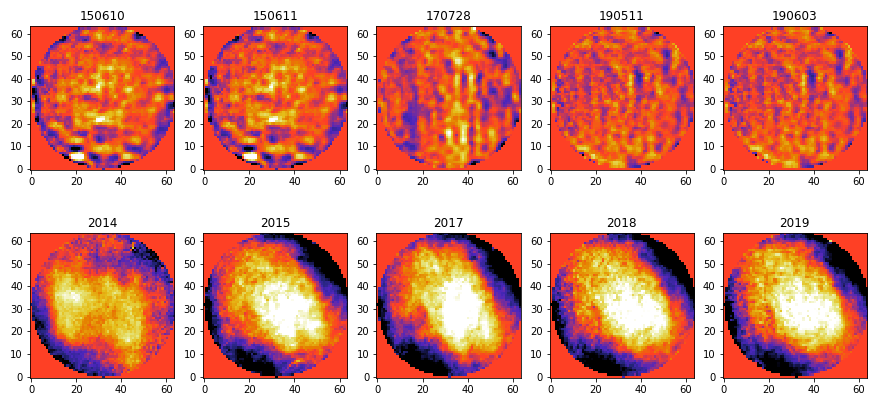}
    \caption{\label{fig:tvar}
    Center (top) and corner (bottom) phase maps at different dates (displayed in YYMMDD format on top of each panel). The corner phase maps have been subtracted by their corresponding central phase map. The color scale is the same as in Figure~\ref{fig:center_to_edge}.}
\end{center}
\end{figure}

The phase maps at the center of the detector, at the position (or close to) where image sharpening is performed, differs from day to day as image sharpening iterations might show slightly different residuals (top panels Figure~\ref{fig:tvar}). 
The daily variation of the central aberrations might be connected to the image sharpening procedure. Image sharpening is done in an iterative way: the deformable mirror has an array of correction voltages applied at each iteration until sufficient convergence has been achieved. 
The voltage to phase shift is not very well know and can drift. 
This phase shift is constant over the field, but it drifts on a daily time scale.  
The RMS of the central phase maps is typically 68~nm (the residuals map between nights has an RMS of 94~nm at maximum in our observations).
This central aberrations are common to the whole field of view.

\begin{figure}[htb]
\begin{center}
\includegraphics[width=17cm]{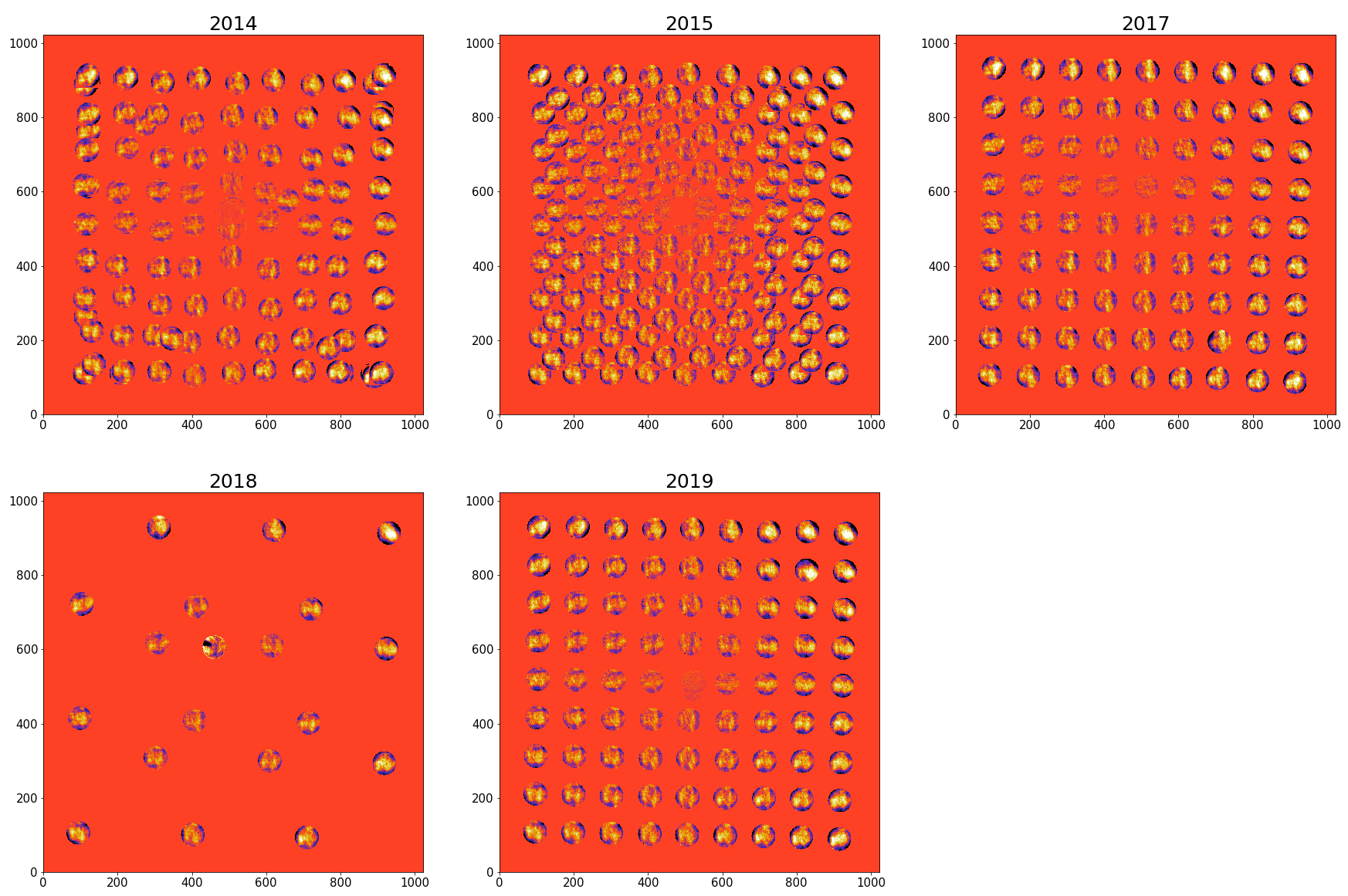}
\caption{\label{fig:grids}
  Center-subtracted phase maps grids gathered in several years across NIRC2 detector. The color scale is the same as in Figure~\ref{fig:center_to_edge}.}
\end{center}
\end{figure}

Gathering phase diversity data across the field of view takes several hours which are typically split into two days or more in order to be consistent with other telescope activities and calibrations.
Therefore, the phase maps derived in different days need to be calibrated by their common global wavefront error, represented by the central phase map.
In practice this corresponds to subtracting from each map the same-day center phase map. 
However, subtracting the center phase map from all other phase maps and feeding them to AIROPA for the instrumental aberrations model 
is mathematically incorrect and does not yield the proper convolution kernel. 
The rigorously correct way to obtain a consistent grid of phase maps is to subtract the center phase map corresponding to each phase diversity data and then add in a central phase map of choice.
To best represent the relevant instrumental aberrations, the central phase map of choice needs to be as close as possible to the science observations to which one wants to apply AIROPA.
To this end, we gather phase diversity data (and derive a phase map) in the center for each science night used in our on-sky testing \cite{Turri22, Terry:inprep}.

\begin{figure}
    \centering
    \includegraphics[width=14cm]{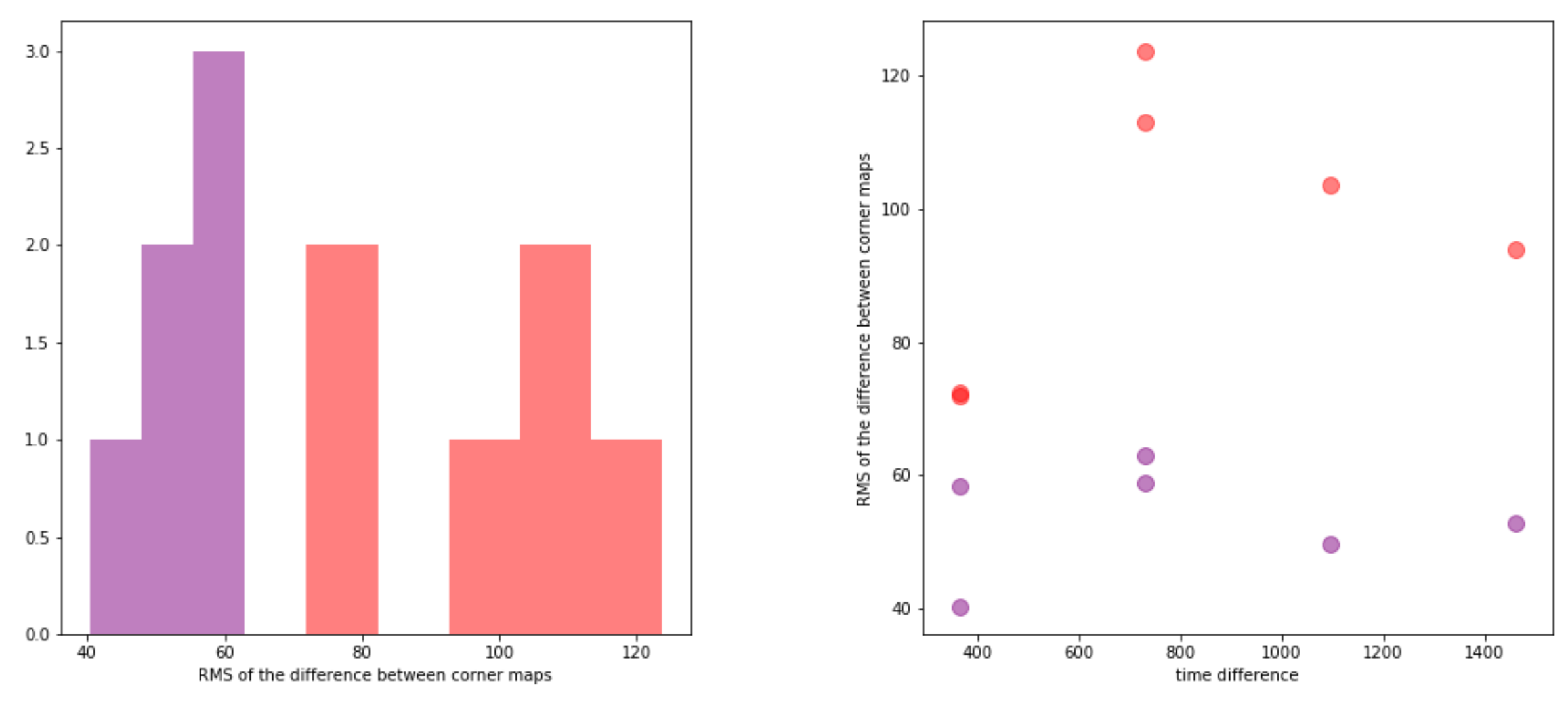}
    \caption{
$Left:$ distribution of the RMS of the difference between upper right corner maps (in nm) for center-subtracted maps (purple) or not (orange). The center-corrected maps show a much more concentrated distribution. $Right:$ same comparison but potted as a function of temporal distance (in MJD).
}
    \label{fig:wfedistr}
\end{figure}

The differential part of the instrumental aberrations can potentially change from year to year due to use and upgrades to the hardware and software. 
To verify the stability of our phase aberrations measurements and to monitor possible changes in the instruments we collected phase diversity data every year since 2014.
The phase maps grids are all reported in Figure~\ref{fig:grids} and differential wavefront error values are reported in Table~\ref{tab:obs}.
One of the big challenges of the project is that the instrument itself has not been built for these calibration procedures. 
We therefore had to investigate and establish a procedure. 
For this reason the data collected throughout the years does not appear uniform, especially in the beginning of the project. 

we find that the differential instrumental aberrations were constant within the measurement uncertainties ($\sim$20 nm).
The average wavefront error at the center position is $\sim$70 nm (based on 16 measurements from 2015 to 2019) and the typical day-to-day variation, which is estimated by taking the RMS of the residual maps between nights, is $\sim$70 nm. 
On the other hand, the differential part of these aberrations remains very stable from year to year across the detector: the RMS of the difference between corner top-right maps (center-subtracted, based on 4 measurements between 2015 and 2019) is about 55 nm (without center-subtraction is 95 nm, see Figures \ref{fig:wfedistr}).

\begin{figure}
    \centering
    \includegraphics[width=14cm]{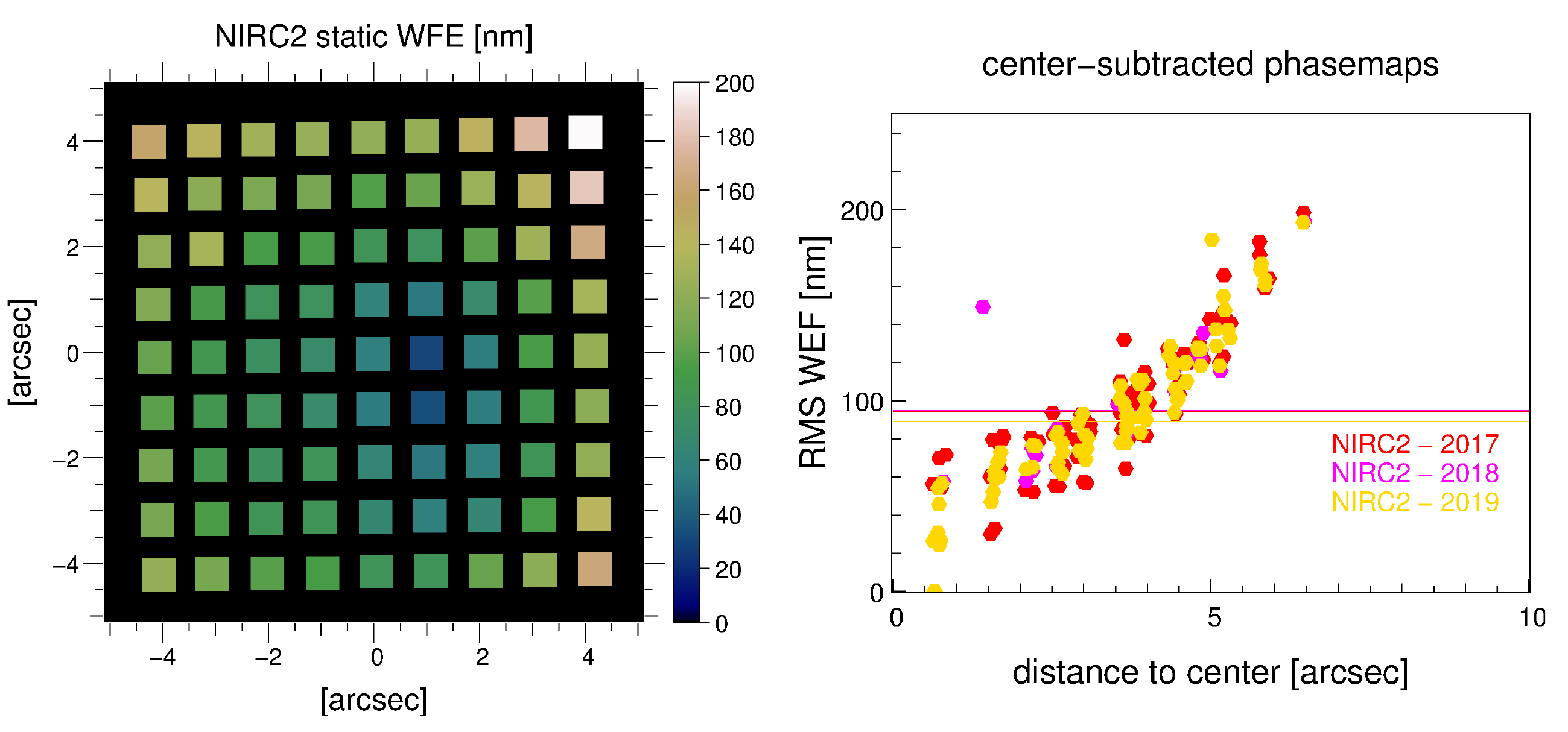}
\caption{
Example of wave-front error (expressed in nm according to the side color-bar) as a function of location across the NIRC2 detector (for 2019). $Right:$ comparison of the instrumental wavefront error as a function of distance from the detector center for several years.}
    \label{fig:acrossdet}
\end{figure}

Considering center-subtracted maps, typically the highest wavefront error (in the to-right corner) is around 200~nm (see Figure \ref{fig:acrossdet}).
In 2014 the phase map in the same area is slightly flatter, with a wavefront error of about 120~nm.
Between our December 2014 and the 2015 phase diversity data the observatory realigned NIRC2's pupil to the telescope's pupil (Rachel Rampy, private communication) by moving the K-mirror that de-rotates the field\cite{Brunelli12}.
This has likely altered the instrumental aberrations. 
Since then no other major change happened in the optical path and the phase maps appear very stable.
The maximum RMS of the difference between 2015 and 2019 years for center-subtracted corner maps is about 59~nm.

Thanks to this comprehensive data we concluded that, once the common variations are removed (i.e. the central phase map is subtracted from all the others) the differential variations across the field of view are very stable (lower panels Figure~\ref{fig:tvar}).
In Section~\ref{subsec:prediction} we further demonstrate that typically these grids are extremely stable from year to year with possibly the only exception is major hardware changes. 

we note that, one important conclusion from this is the necessity of obtaining a central phase map for each day that the position-dependent instrumental aberrations are measured, which can be used with the differential wavefront error to derive position-dependent phase maps across the field of view.

\subsection{Fiber PSF prediction}
\label{subsec:prediction}

\label{sec:results_fiber}
The most basic test of the instrumental model constructed in AIROPA from the phase map grid is a comparison with in-focus fiber images. 
As a test case we use the fiber images and phase maps obtained in 2017.
AIROPA can be run in two ways: 1) in single-PSF mode, where no field dependency of the PSF is taken into account, and 2) variable-PSF mode, where the instrumental and atmospheric models are used to predict the PSF off-axis.
The comparison of these two modes allow us to test the improvements obtained by modeling the PSF variability across the detector.
For this test the atmospheric model is irrelevant and is therefore not taken into account.

\begin{figure}[h]
\centering
\hspace{1cm} SINGLE-PSF  \hspace{5cm} VARIABLE-PSF \\
\includegraphics[width=14cm]{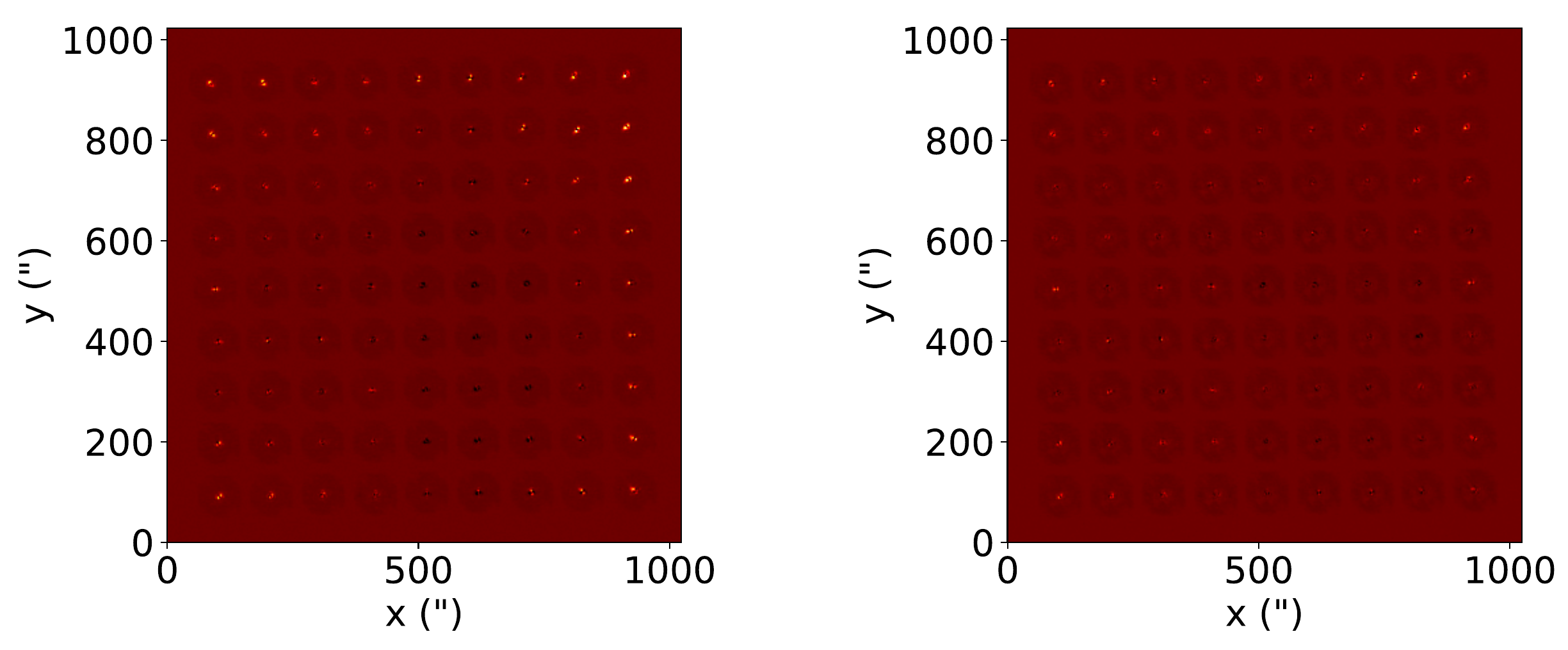}
\caption{Fiber PSF residuals obtained with AIROPA single-PSF (left) and variable-PSF (right) modes for the 2017 data. The color scale is the same in the two panels.
\label{fig:fiber_res}}
\end{figure}

\begin{figure}[h]
\centering
\hspace{1.8cm} RECONSTRUCTED \hspace{0.8cm} RESIDUALS SINGLE-PSF \hspace{0.2cm} RESIDUALS VARIABLE-PSF \\
\vspace{0.5cm}
\includegraphics[width=14cm]{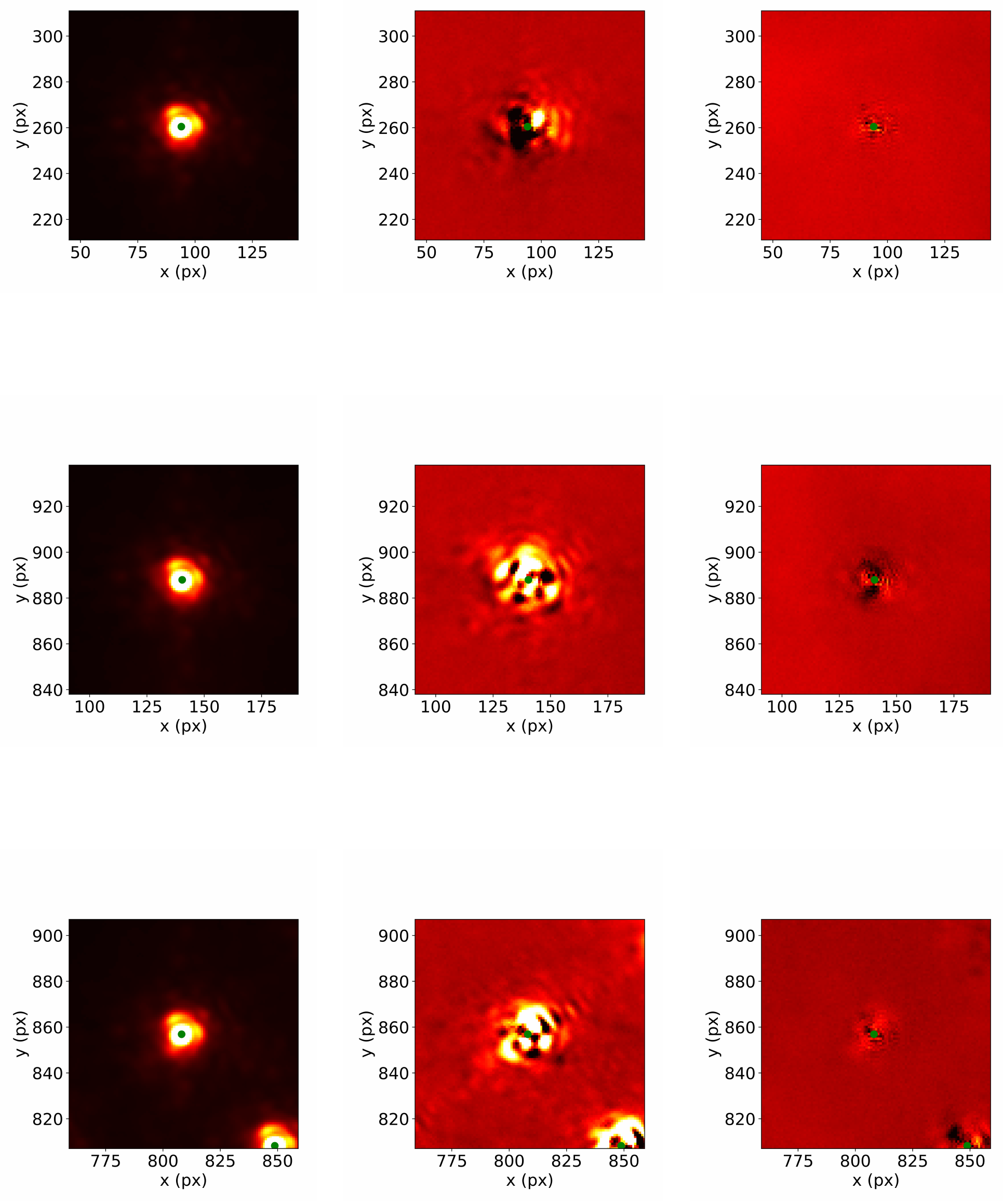} 
\caption{Reconstructed (left) and residual (center and right) images of several fiber images for the 2017 dataset.
The single-PSF and variable-PSF subtracted images have the same color scale.
The green dots mark the center of the detected source.
\label{fig:fiber_res_stars}}
\end{figure}

Figure~\ref{fig:fiber_res} illustrates the residuals of the reconstructed fiber images with respect to the original images for both single- and variable-PSF models.
Figure~\ref{fig:fiber_res_stars} shows a few zoom-in examples of the reconstructed and residuals fiber images using the two AIROPA modes.

\begin{figure}[h]
\centering
\hspace{1cm} SINGLE-PSF  \hspace{5cm} VARIABLE-PSF \\
\includegraphics[width=14cm]{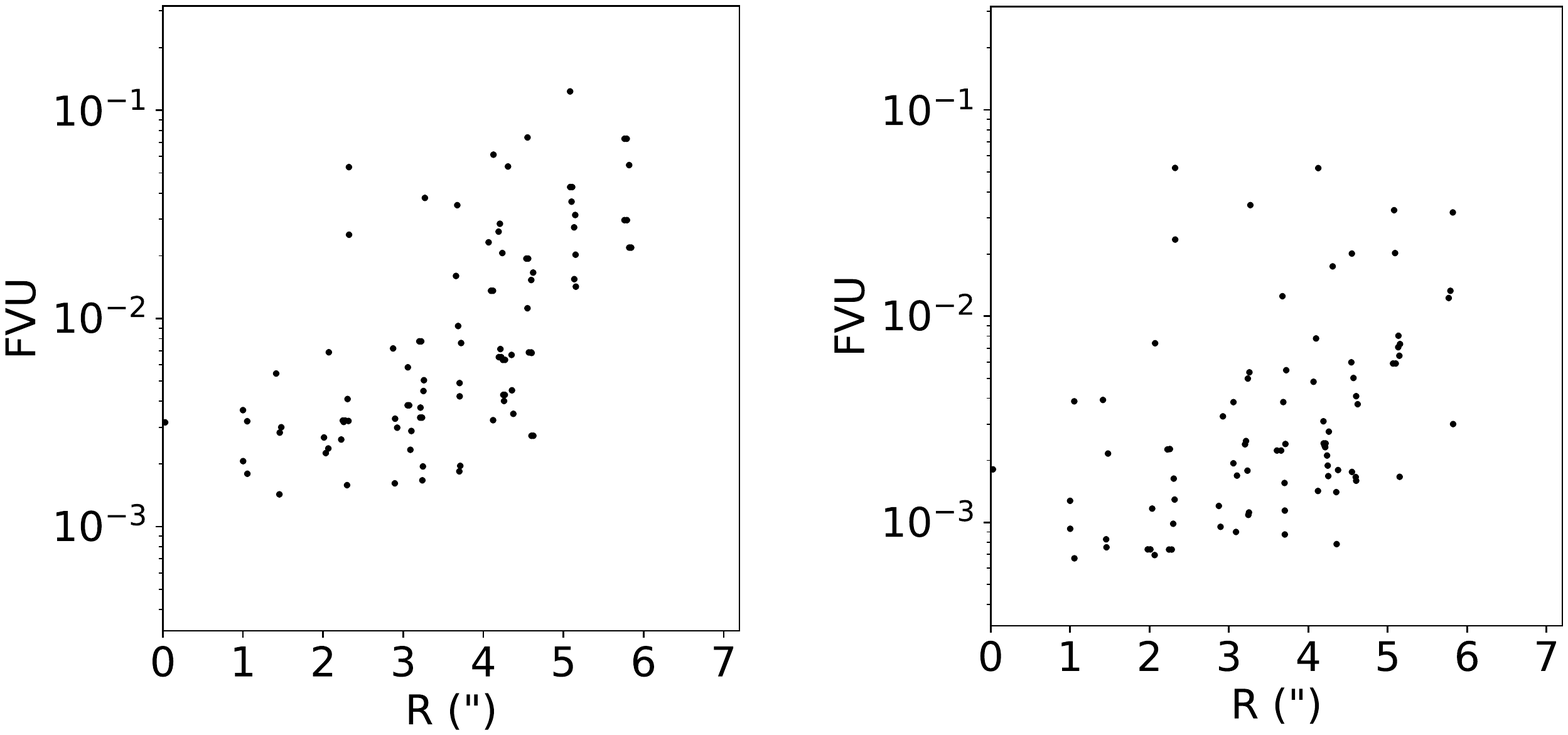}
\caption{FVU of the fiber images as a function of distance from the center of the frame for single-PSF model (left) and variable-PSF model (right).\label{fig:fiber_r}}
\end{figure}

\begin{figure}[h]
\centering
\hspace{1cm} SINGLE-PSF  \hspace{5cm} VARIABLE-PSF \\
\includegraphics[width=14cm]{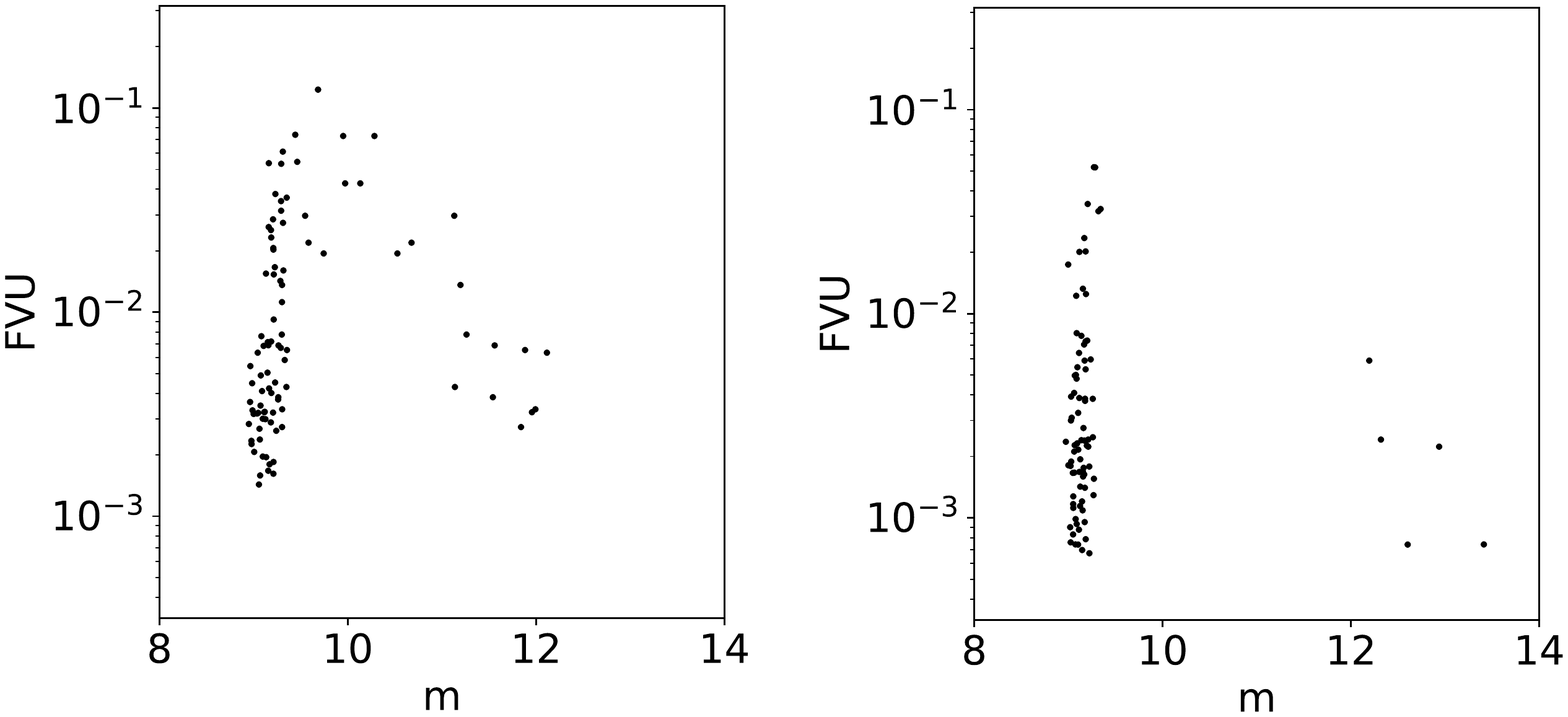} 
\caption{FVU of the fiber images as a function of instrumental magnitude in K. The distribution of points is on average larger for single mode vs. variable mode (which demonstrates variable has better performance).}
\label{fig:fiber_m}
\end{figure}

The variable-PSF model yields much smaller residuals than the single-PSF model.
To quantitatively measure the accuracy of the PSF models, we cannot use the residual positions and magnitudes of the fibers, since their location and intensity on the detector is not known \textit{a priori}. Instead, we have used the FVU: the smaller the FVU, the more accurate the PSF.
In the variable-PSF model, we measure a substantial reduction in the mean FVU of all fibers (from $5.8\cdot10^{-3}$ to $2.4\cdot10^{-3}$), and in its spatial variation (Figure~\ref{fig:fiber_r}).
Moreover, the variable-PSF model shows a better point-source detection as it reduces the number of fake point-source detections (from 15 to 5 out of the 81 knows sources present) caused by speckles in the PSF. 
These are better modelled (and subtracted) when using the variable-PSF model.
This problem is particularly relevant in crowded fields, such as the Galactic Center or Galactic globular clusters, where a bright star with a complex PSF might be mistaken for multiple sources in a field where we already expect several confused sources.
The only features mistakenly detected are the fainter sources: Figure~\ref{fig:fiber_m}) shows that in variable-PSF mode less stars are detected and at lower magnitude with respect to the single-PSF mode. 
Since all fiber source have more or less the same magnitude, the detections at different magnitudes is a proxy for fake source detection.

Furthermore, we used this framework to assess the stability of the instrumental aberrations of the AO system and NIRC2 camera by using the 2017 phase map grid to reconstruct and fit the 2018 in-focus fiber images. 
We find that the mean FVU is reduced by 60\%, (from $4.1\cdot10^{-3}$ to $1.7\cdot10^{-3}$) in variable-PSF mode, similar to what we find by fitting the 2017 fiber images.
This indicates that the aberrations of the instrument are stable on relatively long time scales.
Given the small differences between phase map grids over the years we monitored so far, it's likely that the instrumental aberrations are stable on relatively long time scales.
Variations could happen because of instrument changes and upgrades.
Therefore monitoring the aberrations roughly once a year seems a safe compromise, provided that the global aberrations (residual of the image sharpening) is measured close to the science observations.

We note that the FVU values of both AIROPA modes from the analysis of the 2018 fiber images are lower than those for 2017 data.
This effect is caused by the fact that the 2018 fiber images are more concentrated toward the center of the frame, where residuals are generally smaller.

\section{Conclusions}
\label{sec:conclusion}
We characterized the instrumental aberrations of the NIRC2 imager at W.~M. Keck Observatory with he goal of producing an instrumental model to be used within our off-axis PSF-reconstruction software package AIROPA and monitor the instrument.

We find that the RMS of the wavefront error varies considerably across the field of view, but remains generally stable in time.
What changes more rapidly (even on a daily basis) is the residuals of the image sharpening which produces a wavefront error common to the whole detector.
Therefore, we find that we only need to monitor the differential variations of the aberrations every 1--2 years with a time-consuming grid of positions and we only need to measure the common residuals aberrations in the center of the field (as close as possible in time to science observations).
Regardless, characterizing the instrumental aberrations is time-consuming and a much better approach would be to have the telescope equipped with this type of calibrations capabilities, for example via a pinhole mask that would allow aberration measurements across the field of view in one instance.

We tested our instrumental model with AIROPA by reconstructing fiber images on the detector. 
AIROPA in variable-PSF mode greatly improved the astrometry and photometry accuracy and strongly decrease the number of fake detections.
Even though the instrumental model seems very robust in reproducing calibration data, Ref.~\citenum{Turri22} and \citenum{Terry:inprep} show that the same improvement does not translate onto on-sky data.
Therefore we suspect that the instrumental aberrations that we characterize through afternoon phase diversity are not representative enough of the whole system.
To verify this hypothesis, our next step is to measure aberrations using on-sky phase diversity on bright stars in good atmospheric conditions.

\acknowledgments       
 We acknowledge the support provided by NSF (grants AST-1412615, AST-1518273), Jim and Lori Keir, the Gordon and Betty Moore Foundation, the Heising-Simons Foundation, the W. M. Keck Foundation, Howard and Astrid Preston.
The data presented herein were obtained at the W. M. Keck Observatory, which is operated as a scientific partnership among the California Institute of Technology, the University of California and the National Aeronautics and Space Administration.
The Observatory was made possible by the generous financial support of the W. M. Keck Foundation.
The authors wish to recognize and acknowledge the very significant cultural role and reverence that the summit of Maunakea has always had within the indigenous Hawaiian community.
We are most fortunate to have the opportunity to conduct observations from this mountain.


\bibliography{references} 
\bibliographystyle{spiebib} 


\end{document}